Two-dimensional SDS-PAGE Fractionation of Biological Samples for Biomarker Discovery

Thierry Rabilloud[1,2], Sarah Triboulet[3]

1: CEA Grenoble, iRTSV/LCBM, Chemistry and Biology of Metals Grenoble, France

2: CNRS UMR5249, Chemistry and Biology of Metals, UMR CNRS CEA UJF , Grenoble,  France

3: Université Joseph Fourier, Chemistry and Biology of Metals, Grenoble, France

Correspondence to:

Thierry Rabilloud
iRTSV/LCBM, CEA Grenoble
17 rue des martyrs, F-38054 Grenoble CEDEX 9, France
mail: thierry.rabilloud@cea.fr

Summary
Two-dimensional electrophoresis is still a very valuable tool in proteomics, due to its reproducibility and its ability to analyse complete proteins. However, due to its sensitivity to dynamic range issues, its mots suitable use in the frame of biomarker discovery is not on very complex fluids such as plasma, but rather on more proximal, simpler fluids such as CSF, urine, or secretome samples. Here, we describe the complete workflow for the analysis of such dilute samples by two-dimensional electrophoresis, starting from sample concentration, then the two-dimensional electrophoresis step per se, ending with the protein detection by fluorescence.

1. Introduction

Proteomics in biomarker discovery suffers, maybe more acutely than other areas of proteomics, from two very important caveats. The first one is the necessity of analyzing rather large series of samples in order to take into account the interindividual variablility, and this is a very important requisite for the robustness of the biomarkers. The second caveat lies in the definition of the correct sample. Plasma and serum seem at a first glance as natural choices, as many clinical assays are usually performed on these body fluids. However, the effect of their tremendous complexity on the efficiency of the biomarker discovery process has been very often underestimated, and applying standard proteomic techniques to such complex fluids is generally not an efficient process [1]. Thus, more recent work as focused either on proximal fluids [2] such as CSF [3-7], or urine [8-10] or even samples of secreted proteins obtained from a relevant tissue or cell type model [11-15]. In this case, the implicit rationale is that it will be much easier to find tissue leakage specific markers in the secreted proteins, where they will be much more enriched than in plasma. Then, the presence and specificity of these putative biomarkers can be assessed in the clinically relevant sample (generally a body fluid like serum or plasma) by specific techniques, either antibody-based [16] or mass spectrometry-based [17].

However, from a proteomic point of view, all these samples differ from serum and plasma in the fact that they are much more protein-poor and interfering substances-rich, and there will be a major issue to solve in sample preparation for proteomic studies.

Then, once the sample has been correctly prepared, there is a second crucial choice, i.e. the proteomic platform that will be used for biomarker discovery. Put rather bluntly, there is no discovery proteomic platform today that combines sensitivity and reproducibility. 2D gel-based proteomics is by far the most reproducible platform in discovery proteomics, as exemplified by the much higher demands for gel-based publications in terms of sample numbers as compared to the demands for publishing shotgun proteomics experiments [18]. However, 2D gel-based proteomics lacks sensitivity [19] and ignores some classes of proteins such as membrane proteins [20]. Conversely, shotgun proteomics techniques are much more sensitive and do not present so important protein classes biases. However, they cannot handle easily the large sample series required for biomarker discoveries.

As to the targeted proteomics techniques, such as SRM/MRM, they need first a target to search for, i.e. to have a pre-determined analyte to search for. Moreover, when applied to complex body fluids such as plasma or serum, they are prone to artifacts resulting from spurious degradation of the peptides themselves [21-23].

In this context, 2D gel-based proteomics is still a choice to consider in a biomarker discovery setup, taking advantage of its high reproducibility and its ability to resolve intact proteins. However, because of its rather poor sensitivity, 2D gel-based proteomics must be applied on carefully selected samples (typically not serum or plasma), which must be then carefully handled and concentrated. It is therefore not surprising that one of the very few success stories in proteomic discovered biomarkers has been made by 2D gel electrophoresis on CSF [24], and has led to deep clinical investigation of the serendipitous biomarkers discovered in this way (the 14-3-3 proteins) in several brain pathologies [6], [25-27].
In addition, 2D gel-based proteomics offer unique capabilities when performing reverse probing, i.e. analyzing a patient immunological response to a pathogen to define pathognomonic profiles, as exemplified for candidiosis [28-29].

Thus, this chapter will deal with two different aspects. The first one will be sample preparation, with a focus on sample cleaning and on protein concentration, and the second will be the 2D gel technology itself, with a focus on detection methods.

2. Materials

All solutions are made in ultrapure water (>15 megaohms/cm).

2.1. Protein precipitation/resolubilization

NLS solution: this solution consists of 10% (w/v) N-Lauroyl Sarcosinate in water. Dissolution is best obtained by using hot water. The solution must be kept at room temperature, but must be changed every 2 weeks

TCA solution: this solution consists of 100% (w/v) tricholoroacetic acid in water. This can be either purchased ready-made (as a 6.1M solution of tricholoroacetic acid) or made by adding 450ml of water to a 1kg bottle of tricholoroacetic acid (see Note 1)

Tetrahydrofuran, kept in the refrigerator

Protein solubilization solution: this solution contains 7M urea, 2M thiourea, 4% CHAPS (w/v), 20mM spermine base and 5mM Tris(carboxyethyl)phosphine (see Note 2). This solution is best prepared by sonication in a sonicating bath. It can be kept for months frozen at -20°C and thawed before use

Carrier ampholytes, 3-10 range, sold as a 40% (w/v) solution. Stable for months at +4°C

2.2. Isoelectric focusing

IPG strips. They are availbale from several suppliers and in various pH ranges. Kept frozen at -20°C. For resolution reasons, it is recommended to use long (>15cm) strips.

IPG rehydration solution: this solution contains 7M urea, 2M thiourea, 4% CHAPS (w/v), 0.4% (w/v) carrier ampholytes and 200mM dithiodiethanol (see Note 3). This solution is best prepared by sonication in a sonicating bath. It can be kept for months frozen at -20°C and thawed before use.

IPG rehydration chamber. This is usually a grooved methacrylate block where the widths and lengths of the grooves can accomodate the IPG strips

IPG running device: various chambers exist from different suppliers to run IPG strips.

Some combine the temperature control device, the power supply in a single device, where the rehydration step and the running steps can be carried out sequentially without handling the strips in between. Older devices where the rehydration chamber, running chamber, cooling device and power supply are separated can however be used with equal success, but they usually require some strip handling between rehydration and running

2.3. Equilibration and SDS PAGE

20% SDS solution: prepared by adding 200g of pure SDS to 800ml very hot water (>90°C) under fast magnetic stirring. Is stable for months at room temperature, but SDS may crystallize out if the temperature drops. Should this happen, rewarm at 37°C and shake until redissolved

60%(v/v) glycerol solution. Stable for months at room temperature

Stacking buffer: 120g/l of Tris and 0.8M HCl

Standard gel buffer: 130g/l of Tris and 0.6M HCl (see Note 4)

Alternate gel buffer 1 ( for low molecular weight proteins): 150g/l of Tris and 0.6M HCl

Alternate gel buffer 2 ( for high molecular weight proteins): 110g/l of Tris and 0.6M HCl

Acrylamide solution: 30% w/v acrylamide and 0.8% w/v methylene bis acrylamide. Best to purchase ready made and is stable at +4°C for months.

Ammonium persulfate solution: 10% w/v ammonium persulfate in water. Made fresh every week and kept at room temperature

Temed (tetramethylethylediamine): use as pure liquid. Keep at room temperature and stable for months.

Water-saturated butanol. Mix equal volumes of water and 2-butanol. Shake well and let the phases separate. Keep at room temperature

Multi gel casting chamber. This accessory usually comes as a bundle with multigel running cells (see below)

Multiplate gel running cell. Especially for biomarker discovery, it is recommended to run large series of gels in parallel. Several types of such multi-gel cells are commercially available. They must be coupled to very powerful cooling device and powerful power supplies (not necessary to develop more than 300-500V, but 10-15W/gel are necessary) (see Note 5).

Tank buffer: Tris 6g/l , taurine 25g/l, SDS 1g/l

Equilibration buffer: for 100ml, mix 12.5 ml stacking buffer, 12.5 ml 20% SDS, 50 ml 60% glycerol and 36g urea. Make the day of use or at the earliest the day before. Needs ca. 30 minutes to dissolve at room temperature but must not be warmed for faster dissolution

Sealing agarose: for 100ml, weigh 1g of low-melting agarose and add 12.5 ml stacking buffer, 2ml 20% SDS, and water up to 100 ml. Dissolve by heating to >90°C (e.g. by short pulses in a microwave oven), then add bromophenol blue (0.04% w/v final concentration). Aliquot in 10 ml portions and keep at +4°C.

2.4. Protein detection and image processing

Either commercially available or home made solutions can be used.

Recommended commercial stains are Flamingo from Bio-Rad and Krypton from Thermo-Fisher [30] (see Note 6)

For preparing the home-made fluorescent probe concentrate [31], prepare a solution containing 20mM ruthenium chloride, 60mM bathophenanthroline disulfonate, disodium salt, and 400mM ammonium formate (preferably from a concentrated, titrated solution). Either reflux for 3 days or cook for 3 days in an oven set at 95°C next to a beaker containing water to saturate the oven with water vapor and thus limit the losses in volume by evaporation. The solution should turn very deep orange red. Reconstitute at the initial volume with water, and keep in a bottle at +4°C. Stable for months at this temperature.

For image acquisition, a laser scanner is ideal. Flamingo is best excited at 515nm (emission at 535 nm),  Krypton at  520nm (emission at 580) and the ruthenium complex at 488 nm (emission at 600 nm). If laser scanners are not available, a chamber containing a 302 nm UV table, coupled with either a CCD camera or a simple digital camera and suitable filters (UV + yellow or orange to cut the purple emission of the UV tubes) can also be used. In any case, a spatial resolution of at least 10 pixels/mm is required

For image analysis, a dedicated software is indispensable. Second-generation softwares using image warping as the initial step, are highly recommended.

Fixing solution 1 (Flamingo and Krypton stains): 10% (v/v) acetic acid and 30% (v/v) ethanol. Prepare just before use

Fixing solution 2 (ruthenium complex stain) 1% (v/v) of 85% phosphoric acid and 30% (v/v) ethanol. Prepare just before use.

Staining solution (ruthenium complex stain): 1 µM ruthenium complex in 1% (v/v) of 85% phosphoric acid and 30% (v/v) ethanol. Prepare just before use.

3. Methods

3.1 Sample preparation

The sample preparation described here is convenient for fluid samples (not tissue or cellular samples) that are relatively poor in their protein concentration (down to the microgram/ml range) and rather rich in interfering substances such as salts. It is derived from work carried out on secreted proteins [32]. However, it must be noted that the samples must contain no detergent (Note 7)

If the sample contains more than 0.1M salt, dilute it with water to bring the salt concentration below this threshold. Cool the sample on ice, add lauroyl sarcosinate to 0.1% (w/v) final concentration. Mix well, then add TCA to 7.5% (w/v) final concentration. A voluminous white precipitate forms immediately.

Let the proteins precipitate for 2 hours on ice, then centrifuge at 10,000g at 4°C for 10 minutes. During this time, cool some tetrahydrofuran on ice. Carefully remove the supernatant, then add 1-2ml of cold tetrahydrofuran on the voluminous white pellet. The pellet should literally dissolve in the solvent (or leave very fine particles if the starting amount of proteins is high).

Centrifuge once again at 10,000g at 4°C for 10 minutes. Remove the supernatant. As the pellet is in many occasions completely invisible, it is highly advisable to mark the outside of the tubes in the centrifuge, i.e. where the pellet is. This avoids to aspirate part of the pellet when removing the supernatant.
Repeat the cold tetrahydrofuran wash once. Remove the solvent, then let the tubes dry open at room temperature for 10-15 minutes. (Note 8)

Then add a minimal volume of protein solubilization solution (100-300µl) and let the protein dissolve by sonication for 30 minutes in a sonication bath (Note 9). Recover the solution, and measure the protein content by a Bradford-type assay. Then add carrier ampholytes to 0.4% (w/v) final concentration and store the sample at -20°C.

3.2. Isoelectric focusing

Take the amount of sample required to have 200µg of proteins (as read by a Bradford assay against BSA as a standard). Dilute the sample with IPG rehydration solution to reach the adequate strip rehydration volume (Note 10). If the volume of added IPG rehydration solution represents less than 50% of the final volume, add pure dithiodiethanol to reach a final concentration higher than 100 mM. Add some tracking dye (Note 11) so that the solution is lightly colored.
Add each sample to a groove of the rehydration chamber, then add the IPG strip (Note 12). If necessary, cover with mineral oil to prevent evaporation of the water and urea crystallization. Let rehydrate overnight (Note 13) at room temperature. Then if necessary, transfer the strips in the running chamber. Check the contacts with the electrodes. Run the IPG strips under constant voltage with low voltage initial steps (Note 14). A recommended running program is the following: 100V for 1 hour, then increase from 100V to 300V over 15 minutes, then 300V for 3 hours, then increase from 300 to 1000V over 1 hour, then 1 hour at 1000V, increase to 3500V over 1 hour and 3500 V constant for at least 15 hours (longer times are not detrimental). The movement of the tracking dye becomes usually visible during the 300V plateau, and is more and more visible at the higher voltage phases. No movement of the dye indicates an absence of contact between the IPG gel and the electrodes. The tracking dye should collect at the acidic (anodal) part of the gradient.

If a dyed zone extending beyond the ends of the gradients remains, this is indicative of a high conductivity zone (poor salt removal), and it announces poor resolution 2D gels at the end.

3.3. SDS gel casting

Ideally, the SDS gels must be cast the day before their use to ensure complete and reproducible polymerization. Gel sizes between 150x200 mm and 200x250 mm are recommended. Below the 150x200 size, the 2D gels lack resolution, as the total resolution is proportional to the gel surface in 2D electrophoresis. Above the 200x250 size, the gels become too fragile and too many gels are broken during the final stages of the experiment. The minimal thickness to accomodate an IPG strip is 1mm, but 1.5 mm-thick gels are recommended.

Mount the gels assemblies in the multi gel casting chamber. In order to be able to differentiate each and every gel of the setup, it is recommended to add a distinctive mark cut from thin filter paper at the bottom of each gel assembly, between the glass plates, so that this mark will be embedded in the polyacrylamide gel during polymerization. (Note 15). Separate each gel assembly from its neighbors by inserting a thin polycarbonate plastic foil (usually supplied with the casting chamber) between the gel assemblies. When the gels assemblies are mounted, close the casting chamber.

Many casting chambers allow infusion of the gel solution from the bottom, and this is the recommended procedure to use. This will however require a funnel and silicone tubing to allow casting by gravity. In this case, attach a short piece of silicone tubing (ca 10 cm) to the inlet of the casting chamber, and provide with a stopper. Attach a long piece of silicone tubing (ca 50 cm) to the outlet of the funnel and provide with a stopper. Stoppers that allow easy control of the flow by progressive squeezing of the silicone tube are highly recommended. Then unite the two pieces of tubing with a plastic linker.

The amount of gel mix needed will of course depend on the size of the gels and on the number of gels cast. From the theoretical data (gel dimensions x number of gels), calculate the theoretical volume and add 20% more to take into account the volume that will be lost between and around the gels in the casting chamber. This gives the practical volume needed.

When using the solutions described in section 2, 10%acrylamide gels are easily prepared by mixing 1/6 of the practical volume of concentrated gel buffer, 1/3 of the practical volume of 30% acrylamide solution and 1/2 of the practical volume of distilled water (Note 16). Using the standard gel buffer will lead to a mass window spanning the 15-200 kDa window, with optimal resolution in the 15-40 kDa range and some crowding above. If a higher resolution of the medium and high molecular weight range is desired, use alternate buffer 2. This will lead however to the loss of proteins below 20 kDa. Conversely, if low molecular weight proteins (down to 5kDa) are of interest, use alternate buffer 1. The use of alternate buffer 1 will however increase the crowding in the >30 kDa region of the gel.

When both the gel assemblies and the gel mix are ready, cast the gels. Under slow magnetic stirring (no big whirls) add 0.5 µl of TEMED and 5 µl of 10% ammonium persulfate solution (in this order) per ml of gel mix. Mix for 20 seconds and pour in the casting chamber, avoiding trapping any air bubbles.  Leave a 5mm space at the top of the gels, then overlay each gel with 1ml of water-saturated butanol (upper phase in the bottle). Then close completely the two stoppers, disconnect the linker between the two pieces of tubing, and collect the excess gel mix present in the funnel into a beaker. Clean the funnel and the tubing by running distilled water through them and let dry.
Let the gel polymerize, and check for polymerization in the extra gel mix present in the beaker. When the gels are polymerized, open the casting chamber and recover each gel assembly (plates+ gel). Clean from adhering gel particles under running tap water, rinse with distilled water and pile in a closed box, separating once again the gel assemblies each one from the other by a clean thin polycarbonate foil. These separating foils are cleaned with water and ethanol exactly as the glass plates (see Note 15) and can be reused for months.
Put the closed box containing the gels in the cold room until use. The gels should be used 1-3 days after polymerization.

3.4. IPG strip equilibration, transfer to SDS gel, SDS gel running

At the end of the IPG run, the power is switched off, the paper strips at the ends of the gels (when present) are removed,  and the mineral or silicon oil covering the strips is poured out. It is replaced by at least 10x the strips volume of equilibration buffer. Equilibration is carried out for 20 minutes at room temperature, ideally under reciprocal ("ping-pong") shaking at ca. 45 strokes per minutes. During the equilibration period, the cooling of the second dimension gels is started and the agarose is melted, e.g. in a microwave oven. The agarose should be rather hot when used, to prevent premature setting of the agarose gel.

The second dimension gels are also put out of the cold room, the groove at their tops is carefully dried with a lint free tissue,  and the gels are mounted on their supports to receive the strips.
For each gel, pick a strip with tweezers. X-shaped tweezers, which open only when the fingers exert a pressure on them, are ideal for this purpose. Still holding the strip by one of its ends, insert the other end of the strip in the groove on top of the second dimension gel, pipet 1ml of hot agarose in the groove and complete the transfer of the strip. All air bubbles should be carefully removed before the gel sets. If a side loading of, e.g, a molecular weight standard is desired, a teflon "tooth" is inserted on the side of the second dimension gel at this stage.
Using the paper shapes present at the bottom of each gel, note which strip (and therefore which sample) is loaded on top of which second dimension gel.
When the agarose gel has set, the Teflon tooth can be removed, leaving an empty space to load the control on the side. For example, a 1:1 mixture of sealing agarose and molecular weight standard can be loaded, and this let to set again.
The gels are then mounted in the multi cell, and the tank buffer is poured in the cell. Gels are run at 25V constant voltage for 1 hour, then under constant power up to the end of the run, until the tracking dye reaches the bottom of the gels (Note 17)

3.5. Protein detection

At the end of the run, turn off the electric power and the cooling device. Remove the gels cassettes from the multi gel cell. Wearing gloves, open one gel cassette. Turn the glass plate supporting the gel upside down above the box containing the fixing solution. Fixing solution 1 is used for Flamingo and Krypton stains, and fixing solution 2 for the ruthenium complex stain. Fix the gels for at least 1 hour, then continue with the staining process. Follow the manufacturer's instructions for Flamingo and Krypton stains.

For staining with the ruthenium complex, stain overnight with the staining solution, then destain for 4-6 hours with fixing solution 2. Rinse with water for 5-10 minutes before image acquisition.

Scan with the laser scanner or acquire an image on a UV table with the suitable camera, then process the images with the image analysis software.

Figure 1 shows an example of the results that can be obtained on a secretome sample using the methods described in this chapter.

4. Notes

Note 1: tricholoroacetic acid is highly hygroscopic in its solid state. Thus, is is highly advisable to process a complete and new bottle every time to prepare the 100% TCA solution.

Note 2: This solution provides a basic pH and a high denaturing power, ideal for resolubilizing proteins after precipitation. In addition, it is fully compatible with Bradford-type protein assays.

Note 3: dithiodiethanol provides an almost ideal blocking of thiol groups, without needing any alkylation step [33]. It increases resolution of the basic proteins in the IEF dimensions and simplifies the equilibration process. If desired, thiol alkylation can be performed after protein detection and spot excision.

Note 4: Most of the 2D gels run in the world use the classical glycine-based system [34]. However, we believe that the taurine-based system[35] offers distinct advantages. First, as the gel buffer operates quite close to the pK of Tris, precise control of the pH is easier and offers thus better reproducibility over the long term. Second, it is much easier to tune the resolution jut by changing the pH of the gel buffer, either in the high molecular weight range or in the low molecular weight range [35]. Third, as the ionic strength of the buffer (0.1M) is higher than the one of the Laemmli system, the resolution is slightly higher and the binding of SDS is also higher.

Note 5: Temperature control during SDS PAGE is of great importance to maximize resolution and reproducibility. First of all, the higher the migration speed the higher the resolution, but only if the gels do not heat. Thus, it is very important to evacuate the heat generated during migration by the Joule effect. Due to the geometry of most multi gel cells, this requires both a high cooling power and a powerful pump to overcome the pressure and flow drops in the multi gel cells. As the final result, the optimal migration parameters should be determined empirically and depend both on the geometry of the electrophoresis cell and on the power of the cooling apparatus. Second, it should be kept in mind that the Tris buffers are among those that show the most important changes of pH with the temperature (0.3 pH units/10 °C). Thus, temperature stability from run to run is essential to ensure maximal reproducibility of the migration.

Note 6: In addition to good compatibility with mass spectrometry, important constraints are imposed on the gel detection process. Ideally, the protein detection process should be sensitive, linear over several orders of magnitude to be able to detect changes both for low- and high-abundance proteins, and homogeneous from one protein to another to avoid biases. Recent work has shown that these specifications are best met by fluorescent detection using Flamingo and Krypton stains [30], which operate by environment sensitive fluorescent probes (the probe is not fluorescent in water but fluoresces when bound to proteins).

Note 7: the rationale of the protocol relies on the binding of the lauroyl sarcosinate to the proteins and to the co-precipitation of the proteins and of the lauroyl sarcosinate as its free acid under very acidic conditions. Other detergents (except bile salts) will be soluble under acidic conditions and keep both the lauroyl sarcosinate and the proteins soluble even under acidic conditions. The same solubilizing effect is obtained for samples containing high concentrations of chaotropes (urea, thiourea, guanidine salts). However in this case, efficient precipitation can be obtained if the sample is diluted with water to bring the chaotrope concentration below 1M.

Note 8: in this protein precipitation process, the lauroyl sarcosinate acts as a carrier and has a dual role. Its first role is to carry down the protein precipitate, as proteins do not precipitate well with TCA when they are too dilute (below 1mg/ml) and without carrier. The second role of NLS is to decrease the protein-protein interactions within the pellet, which greatly helps protein resolubilization of the final stage. Tetrahydrofuran has been selected as a washing solvent on a multifactorial rationale. First, it is an excellent solvent of both TCA and NLS in its free acid form. Second, it is a very poor solvent for proteins, so that they do not redissolve prematurely at this stage. Third, tetrahydrofuran is slightly miscible with water, so that traces of the initial supernatant will not form a separate phase during the washing process. This ensures complete removal of the initial acidic aqueous supernatant. Fourth, tetrahydrofuran will be able to remove lipids and other hydrophobic substances that may have coprecipitated with the NLS.

Note 9: A sample preparation protocol based on precipitation and resolubilization requires that both steps are efficient. Thus, as the proteins are very severely precipitated and denatured by TCA, they must be redissolved in highly solubilizing solutions, such as concentrated SDS (for SDS PAGE or shotgun based protocols), or concentrated chaotropic solutions for 2D PAGE-based protocols. Solubilization in intermediate urea concentrations (e.g. 6M urea as in [36]) is inefficient and leads to severe protein losses that are not encountered when the sample is properly resolubilized [32].

Note 10: the optimal final rehydration volume depends on the characteristics of the strips used. It has been experimentally determined [37] that optimal rehydration occurs for final acrylamide concentrations in the rehydrated strip slightly higher than 3%. Most commercial strips are cast as 4% gels with a 0.5 mm thickness. Then the optimal rehydration volume in microliters is given by the formula: strip length (gel part only) in millimeters x strip width in millimeters x 0.65.

Note 11: a tracking dye can be very useful to check for any migration problem. Any anionic dye with no affinity to proteins is suitable. Examples include bromophenol blue, bromocresol green, Orange G, chicago sky blue. The use of several different dyes within a single experiment decreases the probability to change inadvertently the strip order. However, it should be kept in mind that many of these tracking dyes have pH indicator properties. Thus, bromophenol blue, bromocresol green, Orange G take all a similar yellow-orange hue at acidic pH.

Note 12: the ideal pH range for the IPG strip cannot be determined theoretically. It is thus advisable to start with a wide pH gradient (e.g. 3-10 linear) and to adapt the pH range according to the protein density per pH unit. It should also be kept in mind that running purely basic gradients (e.g. 7-10) is more difficult than running gradients covering also an acidic part. Thus, it can be interesting to run each sample twice, on a 3-10 pH gradient to have a good resolution of the basic proteins, and on a 4-7 gradient to have adequate resolution of the acidic proteins, which are in most cases much more numerous than the neutral or basic ones.

Note 13: It has been described that application of a small voltage (50V) during the rehydration step improved protein entry into the pH gradient. This is however possible only when the rehydration and running step take place in the same chamber and thus depends on the apparatus used

Note 14: Low voltage initial steps are required in order to remove smoothly all low molecular weight, charged chemicals (e.g. salts, buffers etc) without generating too much Joule heat that would be detrimental to resolution. The program proposed here has been empirically determined. It should be kept in mind that isoelectric focusing with IPG, on a gradient that is at least 2 pH units wide, requires at least 100Vh/cm2, where the numerator is the integration of volts by time and the denominator the square of the length of the IPG gel.
Some manufacturers recommend to apply, after the cleaning step, a defined power per strip (generally 50µW/strip), as this ensures the highest possible volt.hours in a defined time frame. However, a purely voltage-limited program is safer. If, in a series of samples analyzed in parallel, some are more conductive than others, a collective watt-based program will lead to most of the power passing through the more conductive samples ($P = U^2/R$). Thus, the more conductive samples will dissipate too much power, and consequently too much heat, which will decrease resolution. Moreover, when using a collective watt-based program, the running profile will be different (in volt.hours) from one series of samples to another. This is not the case with a conservative volt-based program as the one described here.

Note 15: the cleanliness of the glass plates is essential to obtain high-quality 2D gels. At the end of each run, remove every gel particle from the glass plates by brushing them under hot tap water. Then rinse each plate in distilled water and let dry. Do not use any detergent to clean the plates, as it is highly likely that some of the detergent will stay on the glass plates despite rinsing and will interfere with the SDS electrophoresis.
Just before use, clean again the glass plates with water and then 95% alcohol, using a lint-free paper tissue. If the plates are really clean, the alcohol rinsing process should produce some wiping noise.
However, with this cleaning scheme, the plates will become dirty over a time frame of several months, and the gel resolution will start to deteriorate. This is easily seen when "tails" begin to appear ahead of the most intense spots. When this happens, clean thoroughly the glass plates with a mildly abrasive dish cleaning powder and a sponge, then rinse profusely under hot tape water, and finally with distilled water.

Note 16: unpolymerized acrylamide is toxic. Wear suitable protection clothes and gloves (preferably powder-free nitrile gloves) when handling gels or acrylamide-containing solutions.

Note 17: the first, initial low voltage step is intended to let the SDS elute completely the proteins from the strip. Once the SDS front has passed the strip, the elution power is minimal, so that vertical trailing is induced. The gels should be then run at maximum speed to limit diffusion, and this maximal speed depends from the gel size, the multi gel cell geometry, and the performances of the cooling system. It must therefore be determined empirically for each system, but manufacturers of multi gel cells generally make useful suggestions in their instructions for use.

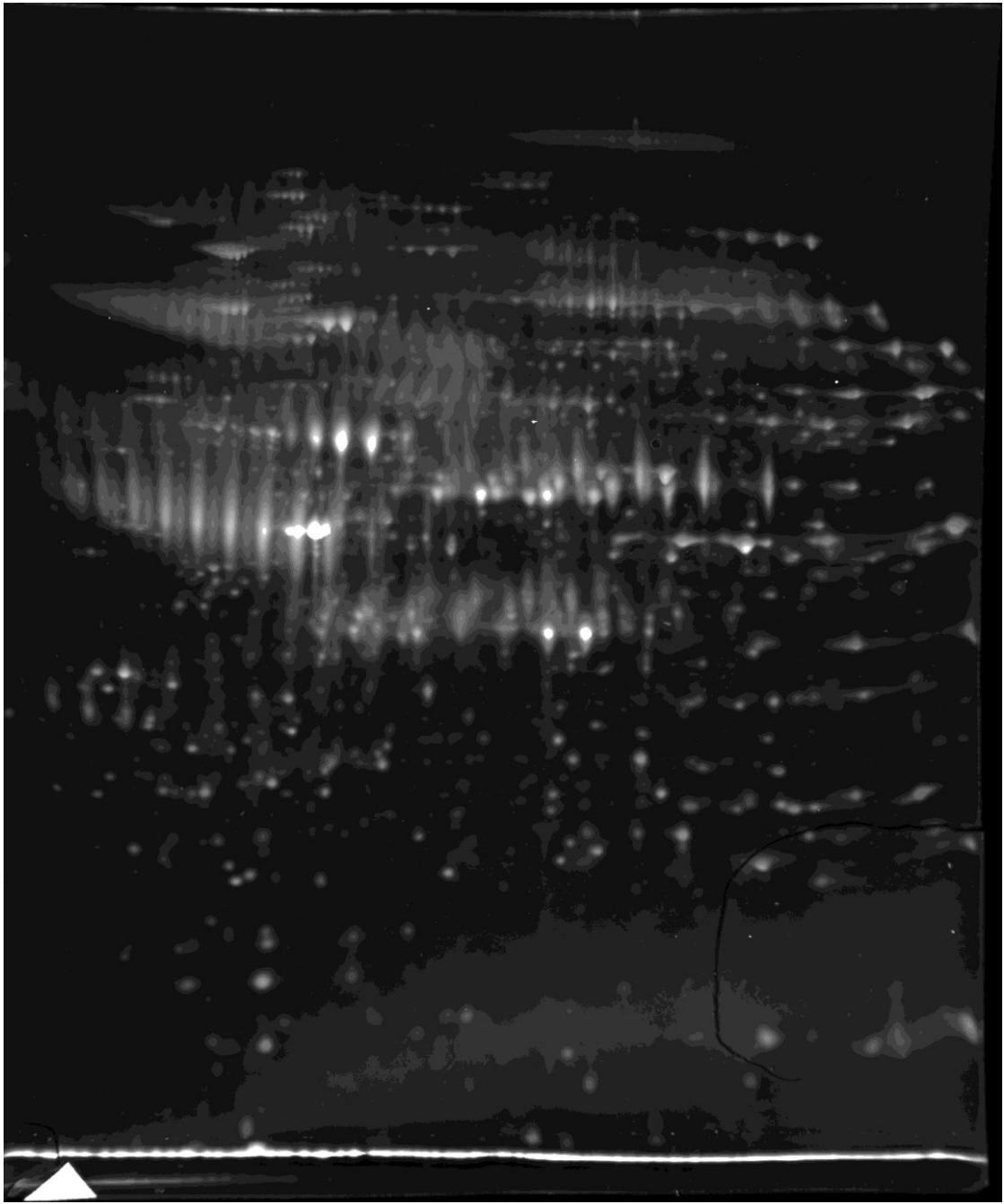

Figure 1:
J774 murine macrophage cells were seeded in a T175 flask (at 100.000 cells/ml) and grown up to confluence in Ultradoma medium supplemented with 1% bovine fetal serum (3 days). The growth medium was then removed, and the cell layer was gently rinsed 3 times with PBS at 37°C, then 3 times with serum-free Ultradoma medium at 37°C. 25ml of serum-free Ultradoma medium were then added and the cells were incubated at 37°C for 24 hours. The conditioned culture medium was then collected, centrifuged for 5 minutes at 1000g to remove floating cells, then centrifuged for 20 minutes at 10,000g to remove smaller debris.
The supernatant was then processed using the methods described in this chapter (TCA-sarkosyl precipitation, analysis by 2D gels, spot detection by fluorescence), to yield the image described in the figure.